\documentclass{aa}  
\usepackage{graphicx}
\usepackage{txfonts}
\begin{document}
%
  %  \headnote{Letter to the Editor}
  \title{Narrow--line AGN in the ISO--2MASS Survey
    \thanks{Based on observations made
      with ESO telescopes at
      La Silla and Paranal (IDs 072.B-0144, 075.A-0345, and
      075.A-0374), with the \mbox{4-m} Blanco telescope at Cerro Tololo
      Inter--American Observatory (CTIO is part of the National Optical Astronomy Observatories, 
      which are operated by the Association of Universities for Research in Astronomy, 
      under contract with the National Science Foundation), 
      with the \mbox{2.2-m} telescope at the
      Centro Astron\'omico Hispano Alem\'an (CAHA) Calar Alto,
      operated jointly by the Max--Planck Institut f\"ur Astronomie and the Instituto de
      Astrof\'isica de Andaluc\'ia (CSIC),
      with the \mbox{1.9-m} telescope at the South African Astrophysical
      Observatory (SAAO), 
      and with the Spitzer Space Telescope, which is operated by
      the Jet Propulsion Laboratory, California Institute of Technology
      under a contract with NASA.}
  }

   \author{C. Leipski
          \inst{1,}\thanks{\emph{Present address:} University of California, Santa Barbara, \mbox{CA--93106}, USA; leipski@physics.ucsb.edu}
          \and
          M. Haas
          \inst{1}
          \and
          H. Meusinger
          \inst{2}
          \and
          R. Siebenmorgen
          \inst{3}
          \and
          R. Chini
          \inst{1}
          \and
	  H. Drass
	  \inst{1}
          \and\\
          M. Albrecht
          \inst{4}
          \and
          B. J. Wilkes
          \inst{5}
          \and
          J. P. Huchra
          \inst{5}
          \and
          S. Ott
          \inst{6}
          \and
          C. Cesarsky
          \inst{3}
          \and
          R. Cutri
          \inst{7}
           }

  % \offprints{Christian Leipski, leipski@physics.ucsb.edu}

   \institute{Astronomisches Institut Ruhr--Universit\"at Bochum (AIRUB),
              Universit\"atsstra{\ss}e 150, 44780 Bochum, Germany
              \and
             Th\"uringer Landessternwarte Tautenburg (TLS), Sternwarte 5,
             07778 Tautenburg, Germany
             \and
             European Southern Observatory (ESO),
             Karl--Schwarzschild--Str. 2, 85748 Garching, Germany
             \and
             Instituto de Astronom\'ia, Universidad Cat\'olica del
             Norte (UCN),  Avenida Angamos 0610, Antofagasta, Chile
             \and
             Harvard--Smithsonian Center for Astrophysics (CfA), 60
              Garden Street, Cambridge, MA--02138, USA
             \and
             HERSCHEL Science Centre, ESA, Noordwijk, PO Box 299, 2200
             AG Noordwijk, The Netherlands
             \and
             IPAC, California Institute of Technology
             (Caltech),  770 South Wilson Avenue, Pasadena, CA--91125, USA
   }

   \date{Received; accepted}

  \abstract
   {A long--standing challenge of observational AGN research is to find
     type 2 quasars, the luminous analogues of Seyfert--2
  galaxies.}
  {We search for luminous narrow--line type 2 AGN, 
  characterise their properties, and compare them with broad--line
  type 1 AGN.}
   {Combining the ISOCAM parallel survey at 6.7$\mu$m with 2MASS,
  we have selected AGN via near--mid--infrared 
  colours caused by the hot
  nuclear dust emission. We performed spectroscopy in the optical
  and, for a subset of the sample, also in the mid--infrared with Spitzer.}
   {We find nine type 2 AGN at redshift $0.1<z<0.5$, three of
  them have even quasar--like [\ion{O}{iii}] luminosities. At the given
  redshift and luminosity range the number of type 2 AGN is
  at least as high as that
  of type 1s. At $z>0.5$ we did not find type 2 AGN, probably
  because the hottest dust emission, still covered by the NIR filters,
  is obscured. The optical spectra of the type 2 host galaxies show
  young and old stellar populations.
  Only one object is an ultraluminous infrared galaxy with starburst.
  The 5--38 $\mu$m spectra
  of the two type 2 sources observed show a
  strong continuum with PAH emission in one case and silicate
  absorption in the other case.}
   {
     %Up to $z\sim0.5$ the number of type--2 AGN is at least as high as
     %that of type--1.
  The near--mid--infrared selection is a successful
  strategy to find luminous type 2 AGN at low $z$. 
  The objects exhibit a large range 
  of properties so that it is difficult to infer details by means of popular SED
  fitting with simple average templates.
   }


   \maketitle
%
%________________________________________________________________

\section{Introduction}

According to the AGN unification scheme type 2 quasars are misaligned type 1s,
so that the central powerhouse is hidden behind a dusty torus seen edge-on
(Antonucci \cite{antonucci93}).  While among radio--loud AGN Spitzer
spectroscopy could show that the powerful FR\,II radio galaxies contain
quasar--like nuclei (Haas et al. \cite{haas05}, Ogle et al. \cite{ogle06}),
the detection of radio--quiet type 2 quasars requires other strategies.  Hard
X--ray surveys turned out to be successful (Norman et al. \cite{norman02}),
but may be hampered by considerable extinction (Vignali et
al. \cite{vignali04}).  In the optical, using the Sloan Digital Sky Survey
(SDSS), type 2 quasar candidates have been found in galaxies with narrow
permitted emission lines and high [\ion{O}{III}]$\lambda$5007 equivalent
widths (Zakamska et al. \cite{zakamska03}) and were confirmed by
spectropolarimetry (Zakamska et al. \cite{zakamska05}).

An alternative approach is to look for the characteristic near-- and
mid--infrared reemission of the hiding dust heated by the strong
radiation field of the AGN.  While several such searches were started
using 3.6\,--\,8.0 $\mu$m data from the Spitzer Space Telescope
(e.g. Lacy et\,al. \cite{lacy04}), at 24\,$\mu$m (e.g. Alonso--Herrero
et al. \cite{alonso06}), and as a combination of 24\,$\mu$m and radio
data (Martinez--Sansigre et al. \cite{martinez05}, Weedman et
al. \cite{weedman06}), we have utilised the ISO--2MASS AGN survey.
The results on the 24 type 1 AGN have been presented by Leipski et
al. (\cite{leipski05}).  Here, we report about the nine type 2 AGN.
We use a $\Lambda$--cosmology with  $H_0=71\,{\rm km\,}{\rm s}^{-1}$,
$\Omega_{{\rm matter}}$ = 0.27, and $\Omega_{\Lambda}$ = 0.73
throughout the paper.

\section{Data}

\begin{figure*}[t!]
  \centering
  \includegraphics[angle=0,width=16cm]{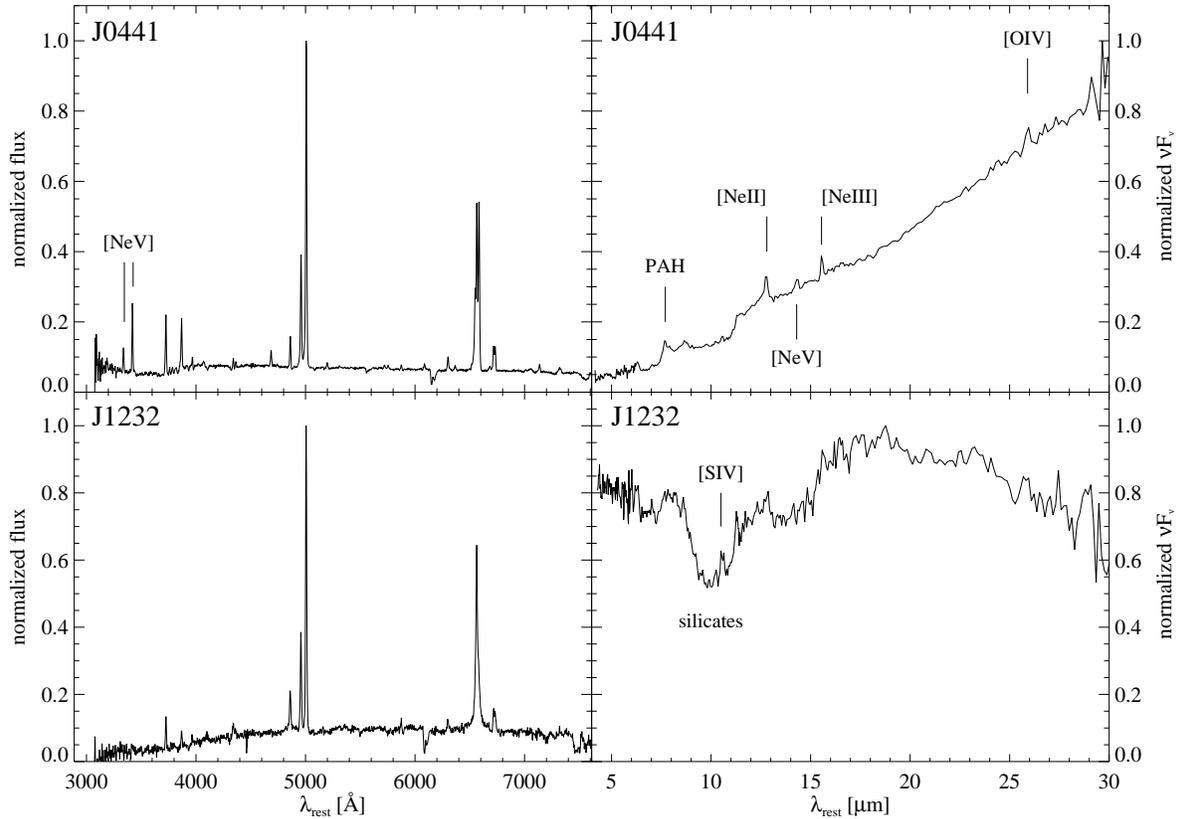}
  \caption{Spectra of two
  ISO--2MASS type 2 AGN. The optical spectra are plotted in
  a linear flux scale, the MIR spectra are shown as
  $\nu\,F_{\nu}$. 
  \label{type2_spectra}
  }
\end{figure*}

We combined the {\it ISOCAM Parallel Survey} at 6.7\,$\mu$m (Cesarsky et
al. \cite{cesarsky96}, Siebenmorgen et al. \cite{siebenmorgen96},  Ott et
al. \cite{ott05}) with 2MASS (Skrutskie et al. \cite{skrutskie06}) in order to
search for AGN largely independent of dust obscuration. At high galactic
latitudes ($|$b$|$$>$20\,deg) the {\it ISOCAM Parallel Survey} covers a total
effective area of $\sim$10\,deg$^{2}$.  We initially considered the $\sim3000$
sources down to F$_{\rm 6.7 \mu m}$ $\sim$ 1\,mJy with MIR, NIR, and optical
photometry available. Then 77 AGN candidates were selected by their red
colours, $H-K_s > 0.5$ and $K_s-LW2(6.7\,{\mu{\rm m}}) > 2.7$ (Vega
system). The colour criteria  were determined by comparison with suitable
samples of various astronomical sources. The details of the ISO--2MASS sample
selection and the begin of the spectroscopic follow--up observations  are
described by Haas et\,al. (\cite{haas04}).  Meanwhile, we completed the
optical spectroscopy for testing the AGN nature. Sources with a Balmer
emission line FWHMs\,$>$\,3000\,km\,s$^{-1}$ are identified as type 1 AGN and
presented in Leipski et al. (\cite{leipski05}).

The optical spectra of the type 2 AGN were obtained at several telescopes
Tab.\,\ref{type2_sources}). The instrument setups have  been chosen to provide
a reasonable combination of spectral resolution  and wavelength coverage.  All
spectra were obtained in long--slit mode using slit widths of
1--1\farcs5\arcsec, depending on the seeing conditions.  The data reduction
and analysis was performed using {\sc Eso/Midas}  version \mbox{04FEBpl1.0}
with standard procedures including bias subtraction, flatfielding, cosmic ray
removal, wavelength and flux calibration.

We have obtained low--resolution 5--38 $\mu$m spectra using IRS (Houck et
al. \cite{houck04}) aboard the Spitzer Space Telescope (Werner et
al. \cite{werner04}) for a random sub--sample of 10 out of the 77 IR-selected
AGN candidates. During completion of the optical spectroscopy of the entire
sample,  two of the sources observed with IRS turned out to be type 2
AGN. Their MIR spectra are presented in this paper.  The IRS integration times
were 2$\times$60s in SL1 and SL2 and 2$\times$120s in LL1 and LL2. The two
different nod positions were subtracted to remove the background and the
resulting frames  were averaged. The averaged spectra were extracted and
calibrated using {\sc Spice} with standard procedures  and the orders were
merged using custom IDL procedures. During this process no significant offsets
between the different orders were recognized.

\begin{table*}
\begin{minipage}[t]{\textwidth}
\caption{Parameters of the ISO--2MASS type 2 AGN. NIR magnitudes are
  taken from the 2MASS PSC (Skrutskie et al. \cite{skrutskie06}).
  We used the following spectrographs: 
  FORS1 with grism GRIS\_300V at ESO--VLT, 
  EMMI with grism Gr\#2 at ESO--NTT,
  the \mbox{R--C Spectrograph} with grating KPGL1-1 at the CTIO
  Blanco 4-m telescope,  CAFOS with 
  grism G200 at the CAHA 2.2-m telescope, and the 
  grating spectrograph with GR\#8 at the SAAO 1.9-m telescope.
}  
\label{type2_sources}     
\centering
\renewcommand{\footnoterule}{}  % to avoid a line before footnotes
%\vspace*{-2mm}                
\begin{tabular}{lc|cccccc|lc}       
\hline\hline                
2MASS (J2000.0) & redshift & $J$ & $H$ & $K_s$ & F$_{\rm 6.7 \mu m}$ & \multicolumn{1}{c}{$L_{[\ion{O}{iii}]}$}  & \multicolumn{1}{c|}{$L_{[\ion{O}{ii}]}$}   & telescope  & nominal resolution\\
                &          & mag & mag & mag   & [mJy]               & \multicolumn{1}{c}{[$10^{8}\,L_{\odot}$]} & \multicolumn{1}{c|}{[$10^{8}\,L_{\odot}$]} &            & \AA/px           \\ 
\hline                                         
J01520465$+$2232015     & 0.113 &    16.613               &                  16.159 & 15.642 & 0.89  &                  0.13 & --   & CAHA & 4.59 \\ % 11683 
J02251432$-$2437154     & 0.104 &    15.790               &                  15.125 & 14.519 & 2.92  &                  0.35 & --   & SAAO & 2.30 \\ % 08080 
J04411405$-$3734369\footnote{Detected with IRAS: F$_{\rm 25 \mu m}$=100\,mJy, F$_{\rm 60 \mu m}$=950\,mJy,F$_{\rm 100 \mu m}$=1600\,mJy.} & 0.236 &    16.629               &                  15.521 & 14.797 & 3.96  &                  4.98 & 0.85 & NTT  & 3.54 \\ % 05855 
J11095861$-$3720374     & 0.173 &    16.521               &                  15.613 & 14.863 & 2.00  &                  0.38 & 0.07 & VLT  & 2.64 \\ % 05898 
J12324114$+$1112587     & 0.249 &    16.772               &                  15.474 & 14.261 & 7.12  &                  1.35 & 0.18 & NTT  & 3.54 \\ % 10981 
J14563296$-$0847490     & 0.079 &    16.460               &                  15.882 & 15.374 & 1.69  &                  0.04 & 0.02 & NTT  & 3.54 \\ % 09463 
J17582331$+$5125419     & 0.201 &    16.742               & \hspace*{-2mm}$>$16.180 & 15.774 & 1.26  &                  0.79 & --   & CAHA & 4.63 \\ % 14577 
J19110553$+$6742507     & 0.171 & \hspace*{-2mm}$>$17.755 &                  15.683 & 14.483 & 3.73  & \hspace*{-1.6mm}11.28 & --   & CAHA & 4.59 \\ % 15908 
J22090602$-$3257505     & 0.425 &    16.701               &                  15.778 & 15.134 & 1.92  & \hspace*{-1.6mm}13.76 & 3.81 & CTIO & 1.21 \\ % 06172 
%                       &       &       &       &      \\
\hline		
\end{tabular}
\end{minipage}
\end{table*}

\section{Results and discussion}

\begin{figure}
 \includegraphics[angle=0,width=8.3cm]{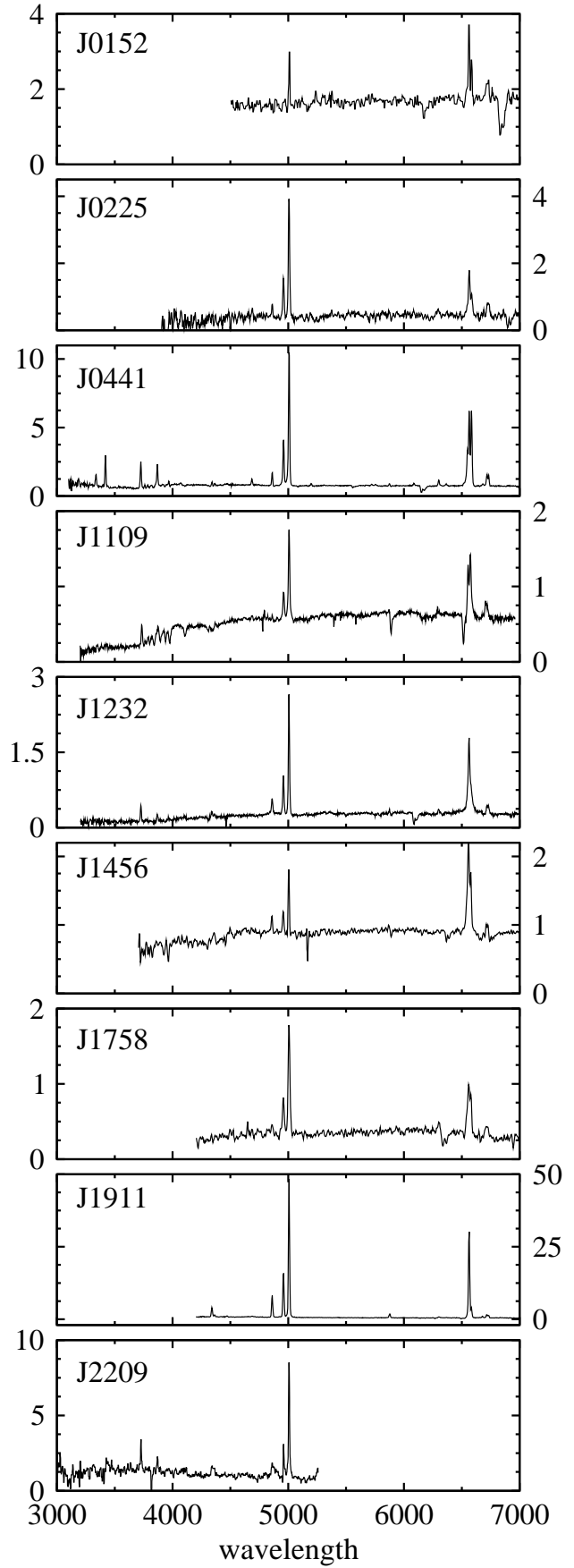}
  \caption{Spectra of the 9 type 2 AGN in the rest frame. 
           All fluxes are shown in $10^{-16}$\,erg\,s$^{-1}$\,cm$^{-2}\,\AA^{-1}$.
  \label{spectra_all}
  }
\end{figure}

Fig.\,\ref{spectra_all} shows the optical spectra of the 9 type 2 AGN.   They
were distinguished from the emission--line galaxies in the diagnostic
line--ratio diagrams (Baldwin et al. \cite{baldwin81}), using the dividing
lines of Kewley et al. (\cite{kewley01}; Tab.\,\ref{flux_ratios}).  This
procedure yields nine type 2 AGN (Tab.\,\ref{type2_sources}). Three of the
type 2 objects can be regarded as type 2 QSOs with [\ion{O}{iii}] luminosities
greater than $3\cdot10^{8}L_{\odot}$ (according to the criterion of Zakamska
et al. \cite{zakamska03}; Tab.\,\ref{type2_sources}). In addition to the red
$H-K_s > 0.5$ sources, the ISO--2MASS sample contains 56 blue sources with
$H-K_s<0.5$ and $K_s-LW2(6.7\,{\mu{\rm m}}) > 2.7$.  For 24 of them we have
obtained spectroscopy (randomly chosen). Only one source (J17582331+5125419)
turns out to be a type 2 AGN. It may actually belong to the red sample, since
it has only an upper limit in $H$ and, thus, a lower limit $H-K_s>0.41$
(Tab.\,\ref{type2_sources}).

None of the sources is listed in NED as X--ray source and only two have
detected radio emission ($\sim$\,1 mJy from FIRST for J12324114$+$1112587 and
38 mJy from NVSS for J11095861$-$3720374).

\begin{table}[b!]
\caption{Diagnostic flux ratios and reddenings calculated from the values given in Tab.\,\ref{type2_fluxes}.}
\label{flux_ratios}
\centering                          % used for centering table
\begin{tabular}{lccc}
\hline\hline  \\[-2.2ex] 
2MASS (J2000.0) & \multicolumn{1}{c}{log$\frac{{\rm [\ion{O}{iii}]}}{{\rm H}\beta}$} & log$\frac{{\rm [\ion{N}{ii}]}}{{\rm H}\alpha}$ & $A_V$  \\ [0.5ex]   % table heading
\hline                      
J01520465$+$2232015$^a$ & 1.38 &              $-$0.24 & $>9.35$         \\  %11683 
J02251432$-$2437154     & 0.96 &              $-$0.28 & $0.61 \pm 0.10$ \\  %08080   
J04411405$-$3734369$^b$ & 1.04 &  \hspace*{1.8mm}0.03 & $<2.76$         \\  %05855
J11095861$-$3720374$^b$ & 1.22 &              $-$0.11 & $<5.05$         \\  %05898  
J12324114$+$1112587     & 0.84 &              $-$0.43 & $2.41 \pm 0.04$ \\  %10981  
J14563296$-$0847490     & 0.55 &              $-$0.19 & $3.12 \pm 0.12$ \\  %09463 
J17582331$+$5125419     & 0.98 &              $-$0.08 & $1.03 \pm 0.25$ \\  %14577  
J19110553$+$6742507     & 0.80 &              $-$0.91 & $1.18 \pm 0.01$ \\  %15908 
J22090602$-$3257505     & 0.85 &  --	              &  --             \\  %06172 
\hline               
\end{tabular}
\begin{list}{}{} 
 \item[$^a$] For this source the H$\beta$ line could not be identified within the noise. We only give an upper limit on the H$\beta$ flux and, thus,  a lower limit on $A_V$.                   
 \item[$^b$] These objects show Balmer absorption. Since we do not correct for the stellar population, the H$\beta$ flux is a lower limit and, thus, the $A_V$ is given as an upper limit.                   
\end{list}
\end{table}

The emission--line fluxes of the type 2 AGN were determined  by fitting
gaussians to major emission lines. We did not correct  the spectra for the
underlying stellar continuum. Thus, for objects  with strong
young-to-intermediate age stellar populations  (J0441, J1109) the fluxes
especially of H$\beta$ are likely to be  underestimated.  When we account for
this effect, it does not change the classification of the objects.  The
dominant source of uncertainty in deriving the line fluxes is the placement of
the continuum next to the emission line. Following Bennert et
al. (\cite{bennert06a,bennert06b}) we therefore estimated the errors of the
flux measurements as the product of the root--mean square deviation of the
local continuum and the FWHM of the line. We assumed a  minimum error in the
flux measurement of at least 5\% though.  The flux values and the
corresponding errors  are given in Tab.\,\ref{type2_fluxes}.

\begin{table*}%[t!]
\begin{minipage}[t]{\textwidth}
\caption{Fluxes of major emission lines. All fluxes are given in $10^{-16}$\,erg\,s$^{-1}$\,cm$^{-2}$. The errors ($\Delta$) are given in \%.}  
\label{type2_fluxes}     
%{\footnotesize
\centering
\renewcommand{\footnoterule}{}  % to avoid a line before footnotes
\begin{tabular}{lcccccccccc}    
\hline\hline       \\[-2.2ex]         
2MASS (J2000.0) & [\ion{O}{ii}] & $\Delta$([\ion{O}{ii}]) & H$\beta$      & $\Delta$(H$\beta$) & [\ion{O}{iii}] & $\Delta$([\ion{O}{iii}]) & H$\alpha$     & $\Delta$(H$\alpha$) & [\ion{N}{ii}] & $\Delta$([\ion{N}{ii}])  \\ 
                & $\lambda3727$ &  \%                     & $\lambda4861$ &      \%            & $\lambda5007$  &  \%                      & $\lambda6563$ &   \%                & $\lambda$6583 & \%                      \\%[0.5ex]
\hline                                         
%object                  
J01520465$+$2232015      & \multicolumn{1}{c}{--}  & \multicolumn{1}{c}{--} & $<$0.66  & \multicolumn{1}{c}{--} & 15.75    & 6 & 25.52                  & 5                      & 14.85                     &  9               \\
J02251432$-$2437154      & \multicolumn{1}{c}{--}  & \multicolumn{1}{c}{--} & 5.42     &   20                   & 49.67    & 5 & 19.83                  & 7                      & 10.39                     & 13               \\
J04411405$-$3734369      & 19.88                   &  5                     & $>$10.58 &    5                   & 117.10   & 5 & 69.01                  & 5                      & 73.43                     &  5               \\
J11095861$-$3720374      & 3.23                    &  6                     & $>$1.07  &   15                   & 17.89    & 5 & $>$12.95               & 5                      & 10.04                     &  5               \\
J12324114$+$1112587      & 3.70                    & 10                     & 4.04     &    6                   & 28.11    & 5 & 24.01                  & 5                      &  8.82                     &  5               \\
J14563296$-$0847490      & 4.73                    & 20                     & 3.15     &   13                   & 11.11    & 5 & 22.65                  & 5                      & 14.51                     &  5               \\
J17582331$+$5125419      & \multicolumn{1}{c}{--}  & \multicolumn{1}{c}{--} & 2.80     &   25                   & 26.55    & 5 & 11.46                  & 7                      &  9.46                     &  8               \\
J19110553$+$6742507      & \multicolumn{1}{c}{--}  & \multicolumn{1}{c}{--} & 85.71    &    5                   & 546.35   & 5 & 365.36                 & 5                      & 44.53                     &  5               \\
J22090602$-$3257505\footnote{While this object lacks measurements of H$\alpha$ and [\ion{N}{ii}] used for the diagnostic line-ratio diagrams, its high [\ion{O}{iii}]/H$\beta$ and [\ion{O}{iii}]/[\ion{O}{ii}] ratios as well as its redshift and [\ion{O}{iii}] luminosity strongly suggests that this objects harbours an AGN.}  & 22.74                   & 13                     & 11.59    &   15                   & 82.14    & 5 & \multicolumn{1}{c}{--} & \multicolumn{1}{c}{--} & \multicolumn{1}{c}{--}   & \multicolumn{1}{c}{--}   \\
\hline		
\end{tabular}
%}
\end{minipage}
\end{table*}	 

\begin{figure}[b]
  \resizebox{\hsize}{!}{\includegraphics[angle=0]{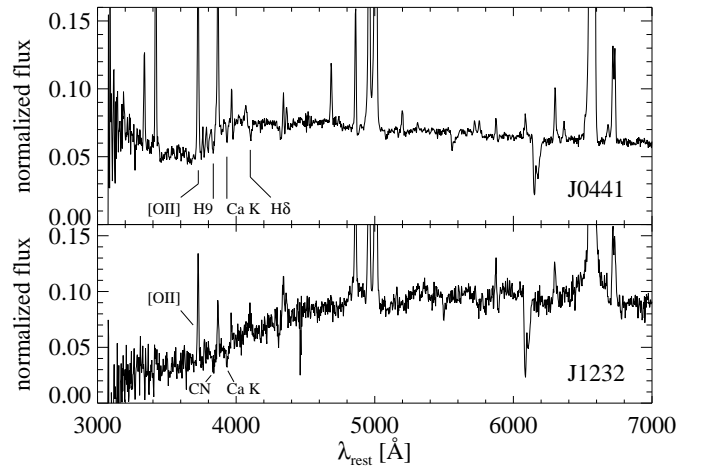}}
  \caption{Zoomed optical spectra of J0441 and J1232. 
  \label{type2_conti}
  }
\end{figure}

\subsection{Type 2 properties}

The ISO--2MASS type 2 AGN show a wide range of properties.  This is not
surprising because according to the unification scheme any details of the host
galaxy can be well discerned in type 2 AGN since the outshining nucleus is
shielded by the dust torus.  In the following we discuss two type 2 objects in
detail for which optical as well as MIR spectroscopy is  available
(J04411405$-$3734369, hereafter called J0441 and J12324114$+$1112587,
hereafter called J1232).  Both share the prominent optical emission lines from
the narrow--line region (NLR), but the signatures of the host galaxies are
rather different (Fig.\,\ref{type2_conti}).

\subsubsection{J0441}
The optical spectrum of J0441 (Fig.\,\ref{type2_spectra}, {\sl upper left})
shows a blue continuum with prominent optical emission lines. The emission
lines include e.g. strong [\ion{Ne}{v}] emission, what is commonly used as a
tracer for an active nucleus. Besides the emission lines, the spectrum of
J0441 clearly shows absorption lines, especially higher order Balmer
absorption (Fig.\,\ref{type2_conti}).  The spectrum indicates a strong
contribution of a moderately young (few hundred megayears) stellar
population. This young population is superposed on an older population
(indicated by the presence of absorption features like \ion{Ca}{II} K;
Fig.\,\ref{type2_conti}).

\begin{figure}[b]
  \resizebox{\hsize}{!}{\includegraphics[angle=0]{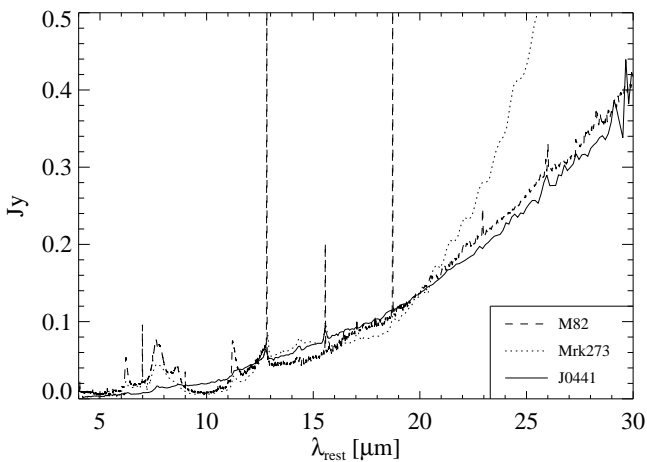}}
  \caption{MIR spectra of J0441, M\,82, and
  Mrk\,273, all scaled to match the flux of J0441 at 20\,$\mu$m. 
  \label{type2_mir}
  }
\end{figure}

That star formation is still ongoing   is supported by the strong
far--infrared (FIR) emission visible in the Spitzer MIR spectrum
(Fig.\,\ref{type2_spectra}, {\sl upper right}). The  star formation is also
indicated by the presence of PAH emission. This emission, although clearly
detected, is of low equivalent width due to a strong NIR and MIR continuum
probably caused by the AGN.  Compared with the starburst galaxy M\,82 (Sloan
et al. \cite{sloan03}), the continuum of J0441 is stronger at 5--20 $\mu$m,
but similar at 20--30 $\mu$m, while equivalent widths of the PAH  features are
much larger in M\,82 (Fig.\,\ref{type2_mir}).  The prominent type 2 AGN/ULIRG
Mrk\,273 has intermediate PAH strength and a very steep slope towards FIR
wavelengths, most likely due to vigorous starbursts
(Fig.\,\ref{type2_mir}). J0441 is the only  ISO--2MASS type 2 AGN which we
detected on IRAS--ADDSCANs  yielding L$_{FIR}$\,$\sim$\,$5\cdot10^{12}\,{\rm
L}_{\odot}$ according to Sanders et al. (\cite{sanders96}, their Tab.\,1).
Thus, J0441 qualifies as a type 2 AGN/ULIRG. Its MIR luminosity at 15\,$\mu$m
of $\nu L_{\nu}(15\,\mu{\rm m})=2.31\cdot10^{45}$\,erg\,s$^{-1}$ places this
object among MIR strong sources that are most likely AGN powered (Ogle et
al. \cite{ogle06}).
  
\begin{figure}[b]
\centering
  \resizebox{\hsize}{!}{\includegraphics[angle=0]{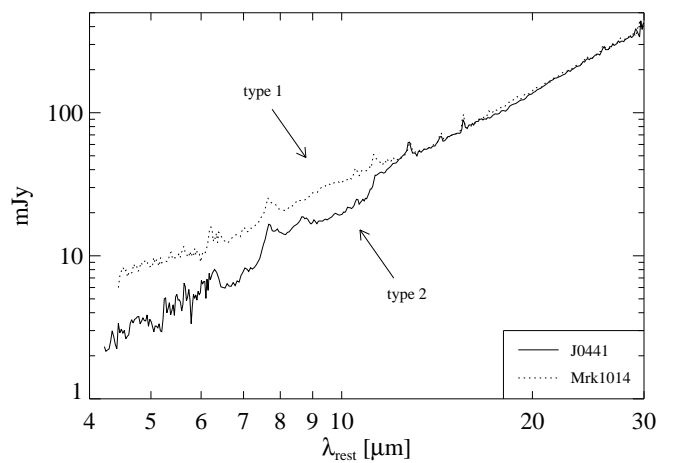}}
  \caption{Comparison of a type 1 AGN with a type 2 AGN in the MIR. 
   Mrk\,1014 was scaled to match the flux of J0441 at 25\,$\mu$m. 
  \label{type1and2}
  }
\end{figure}

Notably, the MIR spectrum resembles that of the FIR luminous dust rich
broad--line quasar Mrk\,1014 (Armus et al. \cite{armus04};
Fig.\,\ref{type1and2}) and possibly is an intermediate case between Mrk\,1014
and Mrk\,273, both showing FIR signatures of dust-enshrouded starbursts
accompanying the AGN to varying degree.  In the comparison with the type 1
source Mrk\,1014 (Fig.\,\ref{type1and2}) the spectra match well at
wavelengths greater than $\sim12\,\mu$m. Below this wavelength the  type 2
source has a significant flux deficit which steadily  increases towards
shorter wavelengths. We will address this issue further in \S 3.2.

Measuring the emission--line fluxes in the MIR spectrum of J0441
(Tab.\,\ref{isocp05855}) and comparing the flux ratios with  those given in
Sturm et al. (\cite{sturm02}) we see that this source falls between AGN and
starburst dominated objects. While it may therefore be classified as a
composite object from the MIR line ratios alone, the optical line ratios, the
[\ion{O}{iii}] luminosity, and the MIR (15\,$\mu$m) luminosity is actually
dominated by the AGN in this type 2 QSO.

\subsubsection{J1232}
In the optical, J1232 displays a red continuum (Fig.\,\ref{type2_spectra},
{\sl lower left}). There are only very few clear absorption lines present with
\ion{Ca}{II} K being the most  prominent (Fig.\,\ref{type2_conti}). No clear
absorption signature can be identified in the spectrum of J1232, neither a
Balmer jump nor a 4000\,\AA~break. This indicates that most stellar features
may have been diluted by a non--stellar continuum.  Since \ion{Ca}{II} K is
visible and higher order Balmer  absorption lines are not, it is unlikely that
the young stellar population is strong in this object.  This is supported by
the flat 15--30 $\mu$m spectrum and the low IRAS--ADDSCAN 3$\sigma$ upper
limit ($F_{60\,\mu{\rm m}}<180$\,mJy), i.e. little  amounts of dust heated by
young stars (Fig.\,\ref{type2_spectra}, {\sl lower right}). The 15\,$\mu$m
luminosity of  $\nu L_{\nu}(15\,\mu{\rm m})=1.06\cdot10^{45}$\,erg\,s$^{-1}$
demonstrates that the MIR emission is powered by a  hidden AGN (Ogle et
al. \cite{ogle06}).

[\ion{Ne}{v}] is not detected in the optical whereas it was even stronger than
[\ion{O}{ii}] in J0441. This can be caused by obscuration of the regions where
[\ion{Ne}{v}] is emitted, i.e. the inner regions of the NLR with the necessary
high--energy photons. In comparison with J0441, the [\ion{Ne}{v}] emitting
region will be more compact because of the less powerful central engine (as
traced by e.g. the MIR luminosity or the [\ion{O}{iii}] luminosity).

\begin{table}[b!]
\caption{Optical and mid--infrared neon fluxes for J04411405$-$3734369. All
  fluxes in $10^{-16}$\,erg\,s$^{-1}$\,cm$^{-2}$ with errors of typically
  $\sim$10\,\%. }
\label{isocp05855}     
\begin{center}                    
\vspace*{-2mm}                
\begin{tabular}{cc|cccc}       
\hline\hline                
[\ion{Ne}{v}] & [\ion{Ne}{iii}] & [\ion{Ne}{ii}]      & [\ion{Ne}{v}]       & [\ion{Ne}{iii}]     & [\ion{Ne}{v}]        \\
$3426$\AA & $3869$\AA   & $12.8\,\mu$m & $14.3\,\mu$m & $15.5\,\mu$m & $24.3\,\mu$m \\
\hline                                         
23.9          & 17.5            & 147.0               & 79.5                & 266.0               & 113.0        \\
%                       &      &        &       &      \\
\hline		
\end{tabular}	
\end{center}
 \emph{Note:}  Flux for [\ion{O}{iv}]\,$\lambda25.9\,\mu$m: 244.0.                   
\end{table}	

Remarkably, we see  broad emission--line components at the bases of H$\alpha$
and H$\beta$ (Fig.\,\ref{type2_conti}), indicating that J1232 has  to be
classified as a type 1.8 according to Osterbrock (\cite{osterbrock77}).  This
also indicates that the central region of the AGN is either viewed under an
intermediate angle by grazing the torus edge and allowing some emission from
the BLR to be visible or that light from  the BLR is scattered into the line
of sight.  If scattered, then we expect to see also the featureless continuum
(FC) from the active nucleus (e.g. Cid Fernandes et al. \cite{cid04}). In this
case, the scattered FC dilutes the absorption lines, especially in the blue
where the FC is stronger and the scattering efficiency is larger. But the
optical continuum of J1232 is red despite the contribution of an intrinsically
blue FC suggesting considerable dust reddening.  This is confirmed by an
emission--line ratio of H$_{\alpha}$/$H_{\beta}\sim6$ corresponding to $A_{V,
NLR}\sim2.4$ (see Tab.\,\ref{flux_ratios}).

\begin{figure}[b!]
  \resizebox{\hsize}{!}{\includegraphics[angle=0]{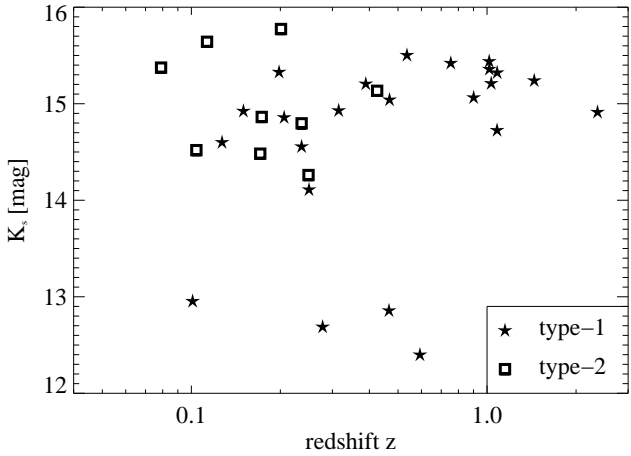}}
  \caption{$K_s$ versus $z$ distribution for the ISO--2MASS AGN.
  \label{k_vs_z}
  }
\end{figure}

\begin{figure}[b!]
  \resizebox{\hsize}{!}{\includegraphics[angle=0]{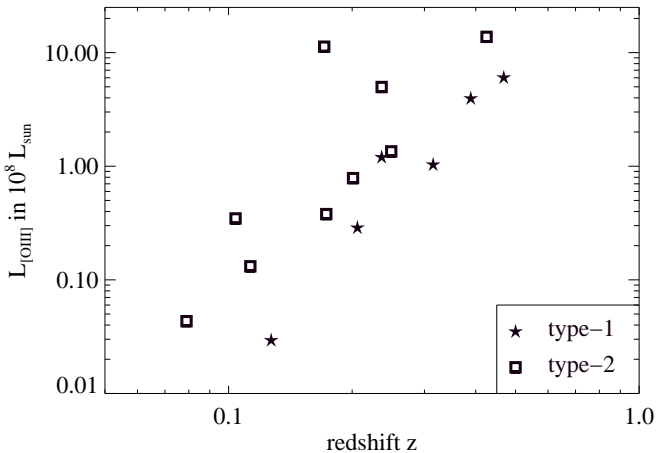}}
  \caption{[\ion{O}{iii}]\,$\lambda$5007 luminosity 
  versus redshift.
  \label{oiii_vs_z}
  }
\end{figure}
\begin{figure}[b!]
  \resizebox{\hsize}{!}{\includegraphics[angle=0]{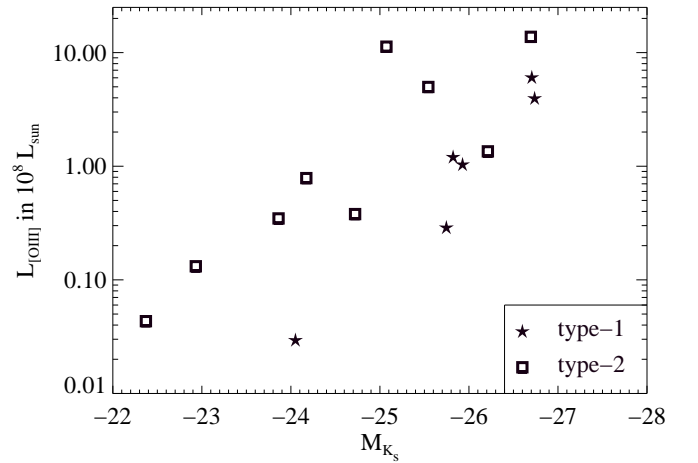}}
  \caption{[\ion{O}{iii}]\,$\lambda$5007 luminosity 
  versus absolute $K_s$ magnitude.
  \label{oiii_vs_k}
  }
\end{figure}

We estimate $A_{V}$ also from the 9.7\,$\mu$m silicate absorption in the MIR
spectra, almost the only clearly discernable feature besides the continuum
(Fig.\,\ref{type2_spectra}).  The estimated depth of the silicate  feature is
sensitive to the placement of the continuum.  Varying this placement allows
us to estimate an uncertainty of the obscuration derived from the silicate
trough. We get a range of $A_{V, MIR}\approx5.7-6.7$  using $A_{V} \approx
17\,A_{9.7\,\mu{\rm m}}$ (Kr\"ugel \cite{kruegel03}) or, alternatively, $A_{V,
MIR}\approx6.2-7.3$  using $A_{V} = 18.5\pm2.0\,A_{9.7\,\mu{\rm m}}$ (Draine
et al. \cite{draine03}).

Since the estimated $A_{V, NLR}$ is only an average value for the NLR, the
difference in the $A_{V}$  values indicates that the absorption for the NLR
gas and the MIR continuum takes place at different regions in the galaxy.
Since most of the NLR emission comes from scales large compared to the MIR
continuum emitter, we suggest that the silicate absorption mainly arises from
dust which is concentrated towards the centre of the galaxy.  The NLR emission
and the MIR continuum is additionally absorbed by ambient dust in the host
galaxy. Such a dust distribution fits well with our findings that the emission
of highly ionised gas in the inner parts of the NLR is significantly obscured by dust.

\subsection{Comparison of type 1 and type 2 objects}
For surveys relying on isotropic properties, e.g. using 178 MHz radio fluxes
(Laing et al. \cite{laing83}, Spinrad et al. \cite{spinrad85}) or FIR data
(Keel et al. \cite{keel05}), the number ratio of type 1 to type 2 AGN turns
out to be $\sim 1$. Surveys in the MIR using the four Spitzer IRAC filters
find also a ratio of $\sim 1$ (e.g. Lacy et al. \cite{lacy07}).  At redshift
$z < 0.5$ we find 12 and 9 type 1 and type 2 ISO-2MASS-AGN, respectively
(Fig.\,\ref{k_vs_z}).  At higher redshift ($z > 0.5$) in fact no narrow--line
ISO-2MASS-AGN has been found, but 12 type 1 AGN. The deficit of type 2 AGN
suggests that, in contrast to the original expectations, our near-mid-infrared
AGN selection strategy is more affected by extinction and hence not isotropic.

To understand this behaviour in more detail, we first consider the origin of
the different parts of the SED of type 2 objects. The dust emission  seen in
the NIR most likely originates at the inner surface of the obscuring torus
that is heated by the central engine.  In type 2 objects the torus is thought
to be inclined such that the central engine is hidden from our direct view.
While at $z<0.5$ a sufficient amount of warm--to--hot dust ($T \sim 1000\,$K)
may be visible, at $z > 0.5$ only the very hottest ($T \sim 1500\,$K) and
inner--most dust is observable in the NIR filters.  The extinction of the most
central and hottest dust emission by the torus or the torus edge leads to a
loss in magnitude and the source  shifts below our magnitude cutoff ($K_s \sim
15.5$\,mag). Also, the relative contribution of the host galaxy  increases at
shorter wavelengths, leading to a bluer $H-K_s$ colour for a redshifted
source.  Since none of the 24 spectroscopically studied ISO--2MASS sources
with $H-K_s<0.5$ and proper detection in $H$ and $K_s$ is a type 2 AGN, we
conclude that the NIR magnitude limit is the main reason for the lack of
high--$z$ type 2 AGN in the ISO--2MASS sample.

If the hottest dust emission in type 2s is affected by obscuration, then such
AGN have to be intrinsically more luminous than type 1 objects to be selected
by our survey.  We test this picture by assuming that the luminosity of the
[\ion{O}{iii}] emission line is a good measure for the total luminosity of the
central engine. This was possible for the nine type 2s and for six type 1s
with $z<0.5$ and $K_s>14$\,mag. The [\ion{O}{iii}] luminosity distribution
shows no difference between type 1 and type 2 AGN (Figs.\,\ref{oiii_vs_z} and
\ref{oiii_vs_k}). However, from Fig.\,\ref{oiii_vs_z} we see that the type 2
sources  have systematically higher [\ion{O}{iii}] luminosities at a given
redshift. If the [\ion{O}{iii}] luminosity is in fact proportional  to the
total luminosity of the AGN, this means that our type 2 sources  are more
luminous than equally distant type 1s.  This is in fact  remarkable if parts
of the NLR emission is indeed obscured as suggested by  Haas et
al. (\cite{haas05}). Then type 2 AGN would tend to show less [\ion{O}{iii}]
than type 1 sources of comparable intrinsic luminosity. This would further
increase the discrepancy  observed here, since we would underestimate the
[\ion{O}{iii}] luminosity and, thus, the total luminosity of our type 2
sources.

In addition, the $K_s$ luminosity at a given L$_{\rm [\ion{O}{iii}]}$ is
systematically lower for type 2 sources than for type 1s
(Fig.\,\ref{oiii_vs_k}).  This supports the idea that type 2 AGN suffer from
obscuration also in the NIR. To compensate for obscured radiation  they need
to have a more luminous central engine to reach our magnitude
limit. Performing a linear least--square fit to either distributions in
Fig.\,\ref{oiii_vs_k} we estimate the difference between the two types of AGN
in M$_{K_s}$ to $\sim$\,$1.3$\,mag at a given [\ion{O}{iii}] luminosity. These
1.3\,mag in $K_s$ corresponds roughly to  $A_{V} \sim 11.6$\,mag (Rieke \& Lebofsky
\cite{rieke85}).

The difference of type 1s and type 2s in the NIR/MIR can also be seen in
Fig.\,\ref{type1and2}. The MIR spectra of a type 1 and a type 2 QSO are
similar at MIR to FIR wavelengths but show an increasing discrepancy for
$\lambda<12\,\mu$m. The type 2 spectrum has a considerable flux deficit
towards the  NIR compared to the type 1 that is assumed to have a largely  unobscured
line-of-sight towards the innermost regions of the AGN.  However, torus models
predict obscuration even at longer MIR  wavelengths (e.g. Pier and Krolik
\cite{pier92}) and observations  indicate that the torus is still affecting
radiation at 24\,$\mu$m  (Shi et al. \cite{shi05}) and becomes optically thin
at $\lambda > 60 \mu$m (Keel et al. \cite{keel94}; Shi et al. \cite{shi05}).
That the type 1 and type 2 MIR spectra look similar even at wavelengths  as
small as 20\,$\mu$m indicates that at these wavelengths the dust is rather
heated by extended starbursts/star formation than coming from a nuclear and
potentially \mbox{(self--)}obscuring torus.   Note that around 18\,$\mu$m the
type 1 object shows additional flux over the type 2 source, likely due to
silicate emission powered by the active nucleus. At shorter MIR wavelengths
the dust emission is powered by the AGN and the obscuration of the torus is
clearly visible. As shown above this trend continues toward the NIR.

\section{Conclusions}
We present a study of NIR/MIR selected type 2 AGN found via the ISO--2MASS AGN
survey.  The optical spectra of the objects reveal different types of host
galaxies.  At least three sources are sufficiently luminous to be classified
as type 2 QSOs. The Spitzer MIR spectra of two sources show very different
features suggestive of a strong starburst contribution in one case and
displaying continuum emission with strong silicate absorption in the other.

In comparison with the ISO-2MASS type 1 sources we find a clear trend for
type 2 sources to be obscured even in the NIR.  This obscuration seems to
continue  into the MIR wavelength range as well.

The flux deficit in the  NIR compared to intrinsically luminous type 1 objects
is identified as the  reason for the lack of type 2 sources at $z>0.5$ in our
survey. This also  results in our detected type 2 sources being intrinsically
more luminous than  type 1 sources at comparable redshifts.

\begin{acknowledgements}
  This work was supported by Sonder\-for\-schungs\-be\-reich SFB\,591 ``Universelles
  Verhalten gleich\-ge\-wichts\-fer\-ner Plas\-men'' der Deutschen
  Forschungsgemeinschaft, and by Nord\-rhein--West\-f\"alische Aka\-demie der
  Wissenschaften.      We are grateful to Vassilis Charmandaris for kindly
  providing us with the IRS spectrum of Mrk\,1014.
  We thank the  anonymous referee for his detailed suggestions.

\end{acknowledgements}


\begin{thebibliography}{}

\bibitem[2006]{alonso06} Alonso--Herrero, A., P\'erez--Gonz\'alez, P.G., 
Alexander, D.M., et al. 2006, \apj, 640, 167
\bibitem[1993]{antonucci93} Antonucci, R. R. J. 1993, \araa, 31, 473
\bibitem[2004]{armus04} Armus, L., Charmandaris, V., Spoon, H.,
  et\,al. 2004, \apjs, 154, 178
\bibitem[1981]{baldwin81}  Baldwin, J. A., Phillips, M. M., Terlevich,
  R., 1981, PASP 93, 5
\bibitem[2006a]{bennert06a}  Bennert, N., Jungwiert, B., Komossa, S., Haas,
  M., Chini, R. 2006, \aap, 446, 919
\bibitem[2006b]{bennert06b}  Bennert, N., Jungwiert, B., Komossa, S., Haas,
  M., Chini, R. 2006, \aap, 456, 953
\bibitem[1996]{cesarsky96} Cesarsky, C. J., Abergel, A., Agnese, P.,
  et\,al. 1996, \aap, 315, L32
\bibitem[2004]{cid04} Cid Fernandes, R., Gu, Q., Melnick, J.,
  et al. 2004, \mnras, 355, 273
\bibitem[2003]{draine03} Draine, B. T. 2003, \araa, 41, 241
\bibitem[2004]{haas04} Haas, M., Siebenmorgen, R., Leipski, C., et
  al. 2004, \aap, 419, L49
\bibitem[2005]{haas05} Haas, M., Siebenmorgen, R., Schulz, B., et al. 2005, \aap, 442, L39
\bibitem[2004]{houck04}
  Houck J.R., Roellig T.L., van Cleve et al. 2004, ApJS, 154, 18
\bibitem[1994]{keel94} Keel, C. W.., de Grijp, M. H. K., Miley, G. K., Zheng, W. 1994, \aap, 283, 791  
\bibitem[2005]{keel05} Keel, W. C., Irby, B. K., May, A., et al. 2005,
  \apjs, 158, 139  
\bibitem[2001]{kewley01} Kewley, L., Dopita, M., Sutherland,
  R., et al. 2001, \apj, 556, 121 
\bibitem[2003]{kruegel03}  Kr\"ugel, E. 2003. The Physics of
  Interstellar Dust (Bristol: Institute of Physics Publishing)  
\bibitem[2004a]{lacy04}  Lacy, M., Storrie--Lombardi, L., Sajina,
  et\,al. 2004, \apjs, 154, 166  
\bibitem[2007]{lacy07} Lacy, M., Petric, A. O., Sajina, A., et al. 2007, \aj, 133, 186
\bibitem[2005]{leipski05} Leipski, C., Haas, M., Meusinger, H., et al.
  2005, \aap, 440, L8 
\bibitem[1983]{laing83} Laing, R. A., Riley, J. M., Longair,
  M. S. 1983, \mnras, 204, 151
\bibitem[2005]{martinez05} Martinez--Sansigre, A., Rawlings, S., Lacy, M., et al. 2005, \nat, 436, 666
 \bibitem[2002]{norman02} Norman, C., Hasinger, G., Giacconi, R.,
  et al. 2002, \apj, 571, 218
\bibitem[2006]{ogle06} Ogle, P., Whysong, D., Antonucci, R. 2006, \apj, 647, 161
\bibitem[1977]{osterbrock77} Osterbrock D. E. 1977, \apj, 215, 733
\bibitem[2006]{ott05} Ott, S., Siebenmorgen, R., Schartel, N., et
  al. 2006, \aap, submitted
\bibitem[1992]{pier92} Pier, E.A. \& Krolik, J.H. 1992, \apj, 401, 99
\bibitem[1985]{rieke85} Rieke, G.H. \& Lebofsky, M.J. 1985, \apj, 288, 618
\bibitem[1996]{sanders96} Sanders, D. B. \& Mirabel, I. F. 1996,
  ARA\&A, 34, 749
\bibitem[2005]{shi05} Shi, Y., Rieke, G.H., Hines, D.C., et al. 2005, \apj, 629, 88
\bibitem[1996]{siebenmorgen96} Siebenmorgen, R., Abergel, A.,
  et\,al. 1996, \aap, 315, L169
\bibitem[2006]{skrutskie06} Skrutskie, M.F., Cutri, R.M., Stiening, R., et
  al. 2006, \aj, 131, 1163
\bibitem[2003]{sloan03} Sloan, G., Kraemer, K., Price, S.,
  Shipman, R. 2003, \apjs, 147, 379
\bibitem[1985]{spinrad85} Spinrad H., et\,al. 1985, \pasp, 97, 932
\bibitem[2002]{sturm02} Sturm, E., Lutz. D., Verma, A., et al. 2002, \aap,
  393, 821 
\bibitem[2004]{vignali04} Vignali, C., Alexander, D., \& Comastri,
  A. 2004, \mnras, 354, 720 
\bibitem[2004]{werner04}
  Werner M., Roellig T., Low F., et al. 2004, ApJS, 154, 1
\bibitem[2006]{weedman06} Weedman, D.W., Le Floc'h, E., Higdon, S.J.U., et al. 2006, \apj, 638, 613  
\bibitem[2003]{zakamska03} Zakamska, N. L., Strauss, M. A., Krolik,
   et al. 2003, \aj, 126,
  2125
\bibitem[2005]{zakamska05} 	
	Zakamska, N. L., Schmidt, G., Smith, P., et al. 2005,
	\aj, 129, 1212 
\end{thebibliography}
\end{document}